# The Asian war bow

Timo A. Nieminen
The University of Queensland, Brisbane QLD 4072, Australia

*Abstract Summary*

*The bow is one of the earliest complex machines, a prime example of the storage and transfer of energy. The physics of the bow illuminates compromises and design choices made in Asian military archery.*

**Keywords- archery; history of technology; military technology**

## I. Introduction

The bow, already developed in the Neolithic, is a complex machine, designed to store energy and deliver it, in a short interval of time, to an arrow (and, subsequently, to a target). The development of the bow from its early beginnings into the various finely engineered forms of medieval and later times demonstrates a deep understanding—even if unarticulated—of the physics governing the bow.

The limitations imposed by real-world materials and economics necessitated compromises in the designs of bows; the requirements of the uses for which they were intended led to further design choices. Investigation of the physics of archery—the behaviour of the complex system composed of the bow, the arrow, the archer, and the target, illuminates these compromises and design choices, and can further our understanding of the role of archery in history.

Military applications of archery, as opposed to hunting and sporting purposes, provide a clear and relatively uniform set of requirements. Thus, we could expect a high level of similarity in war bows over cultures and times and, indeed, effective war bows have many features in common. This is typified by the widespread use of what can be described as a generic Central Asian bow—the Turko–Mongol composite recurve bow [1–4]. This bow is often described as the best bow available before the advent of modern materials and the modern compound bow. However, despite its status as the "best" bow, it was not adopted universally, with limited penetration into Europe, southern India, and South-East Asia—to a large extent, due to the vulnerability of the composite bow to high humidity. More remarkably, following the Manchu conquest of China (1644), the Manchu bow, of similar composite structure, but of very different size, weight, and performance, replaced the Central Asian bow in Chinese military archery.

The decline of the military importance of the bow is often attributed to the adoption of firearms, and, indeed, in Western Europe, the bow was rapidly eclipsed as firearms, especially muskets, improved during the 16$^{th}$ century. For example, in Britain, the longbow, decisive weapon at Agincourt (1415) and Flodden Field (1513), had ceased to exist as a military weapon of any significance by 1622 [5]. Likewise, in Japan, the matchlock musket quickly displaced the bow as the major missile weapon on the battlefield. In China, however, the bow co-existed as a military weapon alongside firearms for almost a millennium [6–8].

The physics of the bow provides understanding of the functional design of the Central Asian composite bow, and illuminates the compromises and choices made in its design. Further insight can be obtained by examining other military bows of significantly different performance. Three such bows that merit such study are the Japanese longbow, or yumi, the Indian steel bow, and, of course, the Manchu bow already noted above. With these examples, we can better understand the interaction between the functional, social, and economic aspects of a technology that has had a major impact on history and the development of the modern world, the Asian war bow.

## II. The bow

A bow consists of two limbs, joined by the grip, by which the archer holds the bow. The bowstring is attached to the limbs near the tips. The side of the bow that faces the archer when in use is the *belly*, and the side away from the archer is the *back*. Bows can be broadly classified by the shape of the limbs. A bow in which the tips of the limbs curve away from the archer (towards the back of the bow) is a *recurve* bow; if the curve is towards the belly, it is a *decurve* bow (a feature of some traditional bows allowing them to remain strung in an unstressed state). If the limb, as a whole, curves towards the back when unstrung, the bow is a *reflex* bow, and if curved towards the belly, *deflex*. In the absence of these types of curvature, the bow is straight.

The Central Asian composite bow is both recurved and reflex, often highly reflexed, with a distinctive "C" shape. In extreme cases, such as Korean bows, the tips of the limbs can overlap [9]. The high level of reflex allows high draw weights and long draw lengths to be achieved using a relatively short bow; this is an important contributor to the efficiency of the design.

However, this design also places extreme demands on the elastic properties of the limbs. As a bow is drawn, the back is placed under an increasing tension, and the belly under an increasing compression. It is difficult to find a single material that will provide sufficient strength under high degrees of both tension and compression, and allow a high degree of deformation. One solution is to use a composite construction, with a belly of horn (e.g., ox horn, buffalo horn, or antelope horn) and a back of a glue-and-sinew composite joined to a central portion of wood. Traditional adhesives that were used included hide glue and fish bladder glue [9]. The construction of a composite bow requires a great deal of skill and time; Turkish bowyers, for example, might wait for up to a year to allow the composite limbs to stabilize before completing the bow [10]. Although composite bows are usually covered by a protective outer layer, they remain susceptible to damage due to high humidity, as the organic glues used will absorb moisture from the atmosphere and weaken [1,10]. This appears to be a key factor that prevented the spread of the composite bow into Europe, southern India, and South-East Asia. While it has also been suggested as a reason for the non-adoption of the composite bow in Japan, it should be noted that the composite bow was used in regions of Korea with a very similar climate, and that the Japanese yumi, of laminated bamboo and wood construction, was also vulnerable to humidity due to the use of similar glues.

The simple alternative to composite construction is the *self bow*, made of a single piece of wood. Self bows are usually straight, and, if powerful, long. The limited ability of woods to survive deformation mean that a self bow must be about 2.3 times as long as the draw length. This requires a long bow—the English longbow was approximately the height of the archer in length, and Japanese longbows were often over 2m in length. In comparison, the Central Asian composite bow is approximately 110cm long on average. Korean bows, while the shortest of the powerful composite bows, under 1m, also had very long draw lengths, with the archer drawing the string back to the ear.

As a device for storing elastic energy, and imparting this energy to an arrow, the performance of a bow depends largely on the stored energy, and the fraction of this energy that can be transferred to the arrow. A further important element of the performance is the velocity of the arrow, which affects the range, the flatness of the trajectory, and the accuracy.

The stored energy is, ignoring losses in the material, equal to the work done in drawing the bow. If the force required to draw the bow a distance of $x$ is $F(x)$, the stored energy is

$$E = \int_{x=0}^{x=D} F(x)dx, \quad (1)$$

where $D$ is the draw length, and $F(x=D)$ is the draw weight. Thus, more energy is stored if the bow has a higher draw weight, and a longer draw length. The maximum draw length is limited by the length of the archer's arms, and the maximum draw weight is limited by the strength and technique of the archer. Training will improve both strength and technique, and can allow a bow of much greater draw weight to be used by a trained archer. A further improvement can be obtained if $F(x)$ is a convex curve, which results in a larger area under the curve. The recurve–reflex design achieves this. (This is also the key improvement in the modern compound bow, which makes use of a pulley system to obtain a curve $F(x)$ for which the force $F$ diminishes once the bow is drawn past a certain point.)

The second factor is the fraction of the energy that can be transferred to the arrow. This is highly dependent on the mass of the limbs of the bow, and the mass of the arrow. At the moment when the arrow leaves the bow, the limbs of the bow are also moving; thus, some of the stored energy is transferred to the limbs of the bow. The center of mass of each limb will be moving at some fraction $f$ of the arrow speed. Therefore, for an arrow of mass $m$, and a bow with limbs of mass $M$, the kinetic energy of the arrow will be

$$E = \frac{F(D)D}{2\left(1 + f^2 \frac{M}{m}\right)}. \quad (2)$$

Thus, a heavier arrow results in more of the available energy being transferred to it. However, the speed of the heavier arrow as it leaves the bow is lower. The dependence of arrow energy and speed on the mass is shown in figure 1.

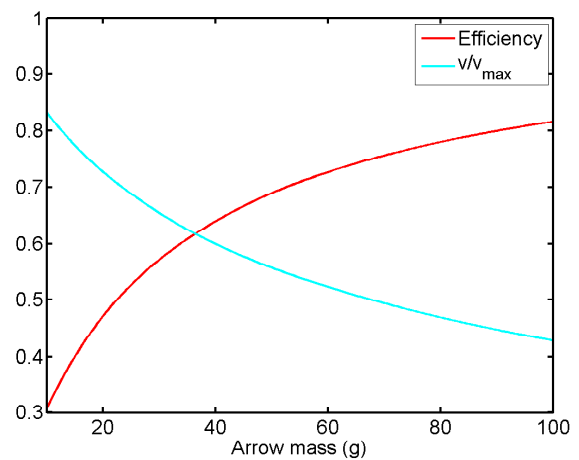

Figure 1. Energy and speed of an arrow, as a function of the arrow mass.

The maximum attainable speed is limited by the mass of the limbs. Ottoman war arrows, shot from the light-limbed Asian composite bow, appear to have varied in mass from 20g to about 40g [1]. In contrast, English longbow arrows appear to have varied from 70–90g. The variation in Ottoman arrow weight at least partly results from the deliberate choice of a range of weights to obtain high speeds for long-range archery (flight arrows), and higher energy for short range use, especially to enhance armour penetration.

### III. THE BOW AND FIREARMS

Firearms have a long history in China, possibly being used before AD 1000 (an early gun, wielded by a demon, is depicted in artwork probably dating from the late 900s [8]), and certainly in use by the 13$^{th}$ century. Eight hundred years later, the bow and firearms were both still in military use.

A key advantage of firearms is the very high projectile energies that are available. An extremely powerful bow is required to achieve 100J of arrow energy, but primitive firearms can exceed this greatly. For example, matchlock muskets can have muzzle energies of 2–3kJ, with projectile energies still remaining over 1kJ at 100m [12]. Muzzle-loading pistols can readily approach 1kJ in muzzle energy, and deliver over 600J at 30m [12]. The thickness of armour required to resist penetration by such energies is approximately 3mm to resist 1kJ, and over 5mm to resist 2kJ [12]. The rapid increase in thickness of European armour in the 15[th] and 17[th] centuries, and its abandonment when required thicknesses became impractical, as firearms improved and became more widely used, can be readily understood. Guns in use at the siege of Pien in China by the Mongols (modern Khai-fun fu), in 1232, were already described as able to pierce any armour [7].

IV. ARMOUR

Arrows shot from a powerful bow are able to penetrate armour of 1mm thickness, unless hitting at a steep angle. Armour 2mm thick, however, can resist penetration by most arrows, most of the time [11]. Before the widespread use of firearms, and, on some occasions, long after this, the most protective armour tended to be sufficient to protect the torso against high-energy arrows. Since this could be achieved by iron or steel plate of approximately 2–3mm in thickness, it was also possible to provide a high degree of protection to other parts of the body, such as the arms and legs. Arm and leg armour was often much thinner (e.g., approximately 1–2mm), and consequently lighter, but could be penetrated by arrows at short range [13]. However, wounds to the arms and legs were less likely to be fatal, or even incapacitating.

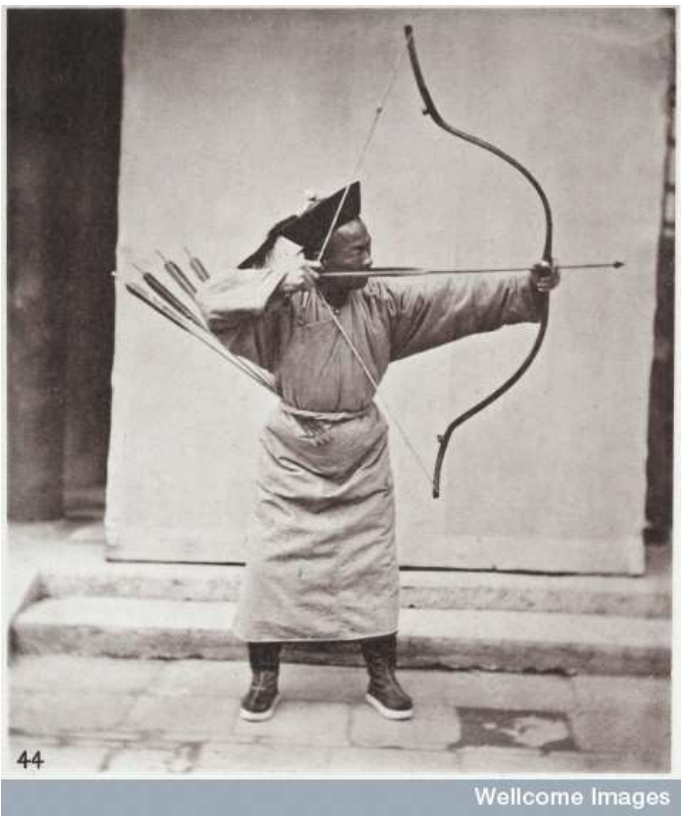

Figure 2. Manchu archer, north China. Photograph by John Thomson 1874, used by permission of Wellcome Images (Wellcome Library, London).

If a large part of the body is armoured, but large parts are not armoured to a standard that could be considered arrow proof, at least at short range, the effectiveness of the bow can be significantly increased by increasing the energy of the arrow. This is a key element of the design of the Japanese yumi and the Manchu bow. Both bows are notable for their large size (a Manchu bow is shown in figure 2).

A large bow can be very powerful, and also allows a long draw length (note the length of the draw in figure 2). This results in the storage of a large amount of energy in the bow. However, a large bow is also a heavy-limbed bow, which limits the speed of the arrow. Consequently, little speed is lost by using very heavy arrows (this results in low-speed arrows, but the maximum possible speed is low, even for very light arrows). Manchu war arrows appear to been approximately 100g on average, and Japanese arrows often exceeded this. One Japanese writer commented that

"For shooting an enemy on the battlefield, one needs, moreover, to practice shooting at a distance of seven or eight ken [approximately 15m] to be able to penetrate his armor. But in tōshiya [a form of sport archery], by sending an arrow light as a hemp stalk a distance of sixty-six ken, how can one hope to pierce armor?" [14]

Both the Japanese and Manchu war bows appear to be highly optimized for close-range penetration of armour. Miyamoto Musashi, in his classic work on swordsmanship, [15] commented that the bow was unsatisfactory if the enemy was more than 40m away. During the Mongol invasions of Japan, it was discovered that Mongol archery could be lethally effective at ranges beyond the maximum possible range of the Japanese archers opposing them. It is noteworthy that the Japanese continued to use their traditional bow even after this rather strong demonstration of the potential of the Central Asian bow.

This mode of archery, focusing on short-range armour penetration, is fulfilling the same role as the handgun. Both the gun and the composite bow are expensive weapons; the gun may well be more suited to mass production given a sufficiently developed iron industry. When advanced matchlock muskets were introduced to Japan, large-scale domestic production was rapidly developed. The English longbow was, in comparison, a much cheaper and easier to make weapon (however, wood of sufficient quality was required, and legal steps were taken to protect domestic supplies and to import suitable wood [5]).

A more important factor when considering the relative merits of the bow and the gun is the training of the soldier using the weapon. As already noted, a high level of training is required for the proper use of powerful bows. For the rapid establishment of large armies, the gun is more useful. When there is a large pool of already trained archers, the higher level of training required by the bow is irrelevant. Chinese governments were able to draw on such a body of trained archers, since the bow was widely used among neighboring populations of nomads, including the Manchu themselves.

The bow, if suitable bows are available, and trained archers are available, does offer many advantages. The rate of fire of a bow is much higher than that of a muzzle-loading firearm, and the ballistics of the arrows compared with a smoothbore musket, coupled with the visibility of the arrow along its trajectory, can allow much more accurate long-range shooting. Fewer specialized materials are required for the manufacture of arrows, compared with the manufacture of gunpowder. (Shortage of gunpowder, in part due to mismanagement and economic difficulties contributed to the defeat of the Ming by the Manchu.)

The gun, however, offers much higher projectile energies. With the advent of modern firearms—breech-loading rifles with metallic cartridges—the balance clearly favours the gun. Prior to that, economic and social factors, especially the training of musketeers as opposed to archers, were more important factors influencing the replacement of the bow by the gun than pure military "effectiveness".

## V. INDIAN STEEL BOWS

An interesting class of bows is the Indian steel bow [3,4]. Steel bows, recurved, but only slightly reflexed, or not at all, were used in regions of India. These bows were typically of only moderate draw weight compared to the most powerful military bows, but were still powerful bows in their own right. They were very heavy-limbed, and unsuited for long-range archery.

What benefits did this design offer? A steel bow is very stable against changes in temperature, high humidity, and so on, which can be mortal enemies of the composite bow. Routine maintenance, largely the prevention of rust, will be sufficient to maintain the bow in working order indefinitely. Long-term storage in arsenals, especially in fortresses, appears to have been the most important factor.

Defensive use in siege warfare makes the short range of the bow less important. An archer shooting from an elevated position within a fortification will gain increased range from this height, and will also gain projectile energy. Very heavy arrows could be, and were, used defensively in sieges (the weight could be sufficient to as to prevent effective use of recovered arrows by the attackers, due to the loss of energy with height).

## VI. CONCLUSION

An understanding of the physics of archery leads to an increased appreciation of the development of the technology of the bow and arrow. Since archery was an important, and sometimes decisive military element, such technological development could have a very broad impact. With the development of the composite bow (by 600 BC or earlier), a highly-optimized design had been achieved; modern materials and modern designs depending on modern materials, such as the compound bow, were required for significant improvement [16].

A variety of tasks led to the development and adoption of highly specialized bows, including variants of the basic composite bow design.

Furthermore, by clarifying the technological and physical factors, the role of social and economic factors can be highlighted.